# Multicast Transmission Prefix and Popularity Aware Interval Caching Based Admission Control Policy

P. Jayarekha

Research Scholar,
Dr. MGR. University, Dept. of ISE, BMSCE,
Bangalore,
jayarekha2001@yahoo.co.in

T.R. Gopalakrishnan Nair

Director, Research Industry and Incubation Center
Dayananda Sagar Institutions, Bangalore
trgnair@ieee.org

***Abstract -*** **Admission control is a key component in multimedia servers, which will allow the resources to be used by the client only when they are available. A problem faced by numerous content serving machines is overload, when there are too many clients who need to be served, the server tends to slow down. An admission control algorithm for a multimedia server is responsible for determining if a new request can be accepted without violating the QoS requirements of the existing requests in the system. By caching and streaming only the data in the interval between two successive requests on the same object, the following request can be serviced directly from the buffer cache without disk operations and within the deadline of the request. An admission control strategy based on Popularity-aware interval caching for Prefix [3] scheme extends the interval caching by considering different popularity of multimedia objects. The method of Prefix caching with multicast transmission of popular objects utilizes the hard disk and network bandwidth efficiently and increases the number of requests being served.**

## 1 INTRODUCTION

Recent advances in high speed networks and communication technologies have made it possible to provide on-line access to a variety of information sources such as reference books, journals, newspapers images, video clips. The two architectures available for servicing the client requests are Client-Pull and Server-Push. In Client-Pull type server streams the data to the client in response to the client's explicit request. Here the client has to determine the playback time, estimate the time to clients request time to frame fetch. While in Server-Push the server serves the client implicitly, in response to the request of the client. The server is responsible for streaming the data in rounds and keeps track of status of each stream. It ensures all the frames are streamed on time within each round.

The challenges faced in designing the multimedia streaming servers is that audio and video requires totally different techniques for their organization and management as compared with numeric data and text. The most critical of these is the continuity requirement. It becomes the responsibility of the multimedia streaming server to ensure that recording and retrieval of media streams to and from disks proceed at real-time rates. Designing a dedicated, multimedia server that is capable of serving multiple clients simultaneously is one of the interesting characteristic features.

Multimedia servers are connected to the clients via ATM (Asynchronous Transfer Mode) networks and its total bandwidth from the storage devices to the server via network to client is fixed. A multimedia storage server can support only a limited number of clients simultaneously. The major concern for a server is to provide service with a good quality to numerous groups of clients keeping the server and network resources in feasible limits. An admission control algorithm determines the acceptance/rejection of a new request. It checks if the available bandwidth is sufficient for total bandwidth required by the streams currently being serviced and the bandwidth requirement of the new request. Based on the desired Quality of Service (QOS), the admission control algorithm decides whether or not to accept the new request. Otherwise, the admission of the new client may introduce distortions or jitter in audio and video quality. Admission control is an integral part of supporting multimedia clients, since it serves as a mechanism to not only identify clients that require multimedia clients that need periodic delivery of a certain amount of data from the disk but also limits the amount of contention for resources like disk that can operate considerably slower than the processor. We are dealing with only clients that are reading from the disk and not writing to the disk. In this paper we have proposed an admission control strategy based on popularity-aware interval caching for prefix, which stores only the prefix of most popular multimedia objects and batches the requests in a batch without the QOS violations of the request and does multicast transmission for all the clients. This reduces the overhead of the hard disk, increases the number of concurrent users and utilizes the network and disk bandwidth efficiently.

The organization of the paper is as follows.

In section 2 we present a background for the VOD Architecture and Multicast and Interval Caching. Section 3 review some existing works on different admission control schemes and multicast, Popularity and Prefix aware interval caching scheme. Section 4 presents Multicast VOD architecture. Section 5 presents Prefix Popularity aware Interval caching Multicast transmission Problem Formulation. Section 6 presents the Proposed Model for Simulation. Section 7 presents simulation model. Section 8 presents simulation results and discussion .In section 9 we conclude the paper and present the future work.

## 2 BACKGROUND

### 2.1 The VOD Architecture

In VOD architecture as shown in the fig 1, multiple users will be able to connect to remote digital video libraries and view videos 'on-demand'. The high speed computer networks will allow users to connect to a massive number of distributed 'video servers', from which users will be able to select and receive high quality video and audio. A VOD system usually consists of several central servers, proxy servers and distributed clients over the entire network. Pre-recorded videos are stored in central servers and sent to clients at their requests through proxy servers. One of the critical computing systems of VOD is the video server. High performance video servers will store a large number of compressed digital videos and allow multiple concurrent clients and proxy servers to connect over the computer network to retrieve a video from the collection of videos. A significant amount of disk and network bandwidth is required for the streaming of multimedia blocks.

The primary requirement of multimedia server is as summarized below

| | |
|---|---|
| **Large Number of Clients:** | A server might have to service thousands of concurrent clients, each requesting for different video or for the same one. |
| **Large Storage Space:** | A collective storage of multimedia files exceeds several terabytes. For example a movie server with a two hundred HDTV quality 20Mbps 2-hour long movie requires nearly 3.6 terabytes of storage. |
| **Large Network and Storage System Bandwidth:** | A movie server that supports one thousand users with HDTV quality movies will require the network and storage bandwidth in excess of 20Gbps. |
| **Real Time Services:** | The services provided to the client should be with a very low latency. |
| **Cost-Effective:** | The architecture must be economically viable. It should be supported with high computing power, in order to support real-time media processing. In addition with providing the storage facility the multimedia server should also support the following resource management functions |
| **File System Support:** | Each server node should perform metadata management, buffer management, cache management. In addition it should provide efficient browsing techniques and content based retrieval. |
| **Admission Control:** | Each storage server keeps track of usage of local resources such as disk bandwidth and network bandwidth and performs local admission control. |
| **Scheduling Support** | Proper replacement technique is also an essential feature of a multimedia server. Scheduling the request and servicing them in a proper sequence is also the responsibility of sever. |
| **Computing Support** | A multimedia server should be embedded with the computing power to perform the functions such as |

transcoding, Speech Recognition, image processing and Character Recognition and so on.

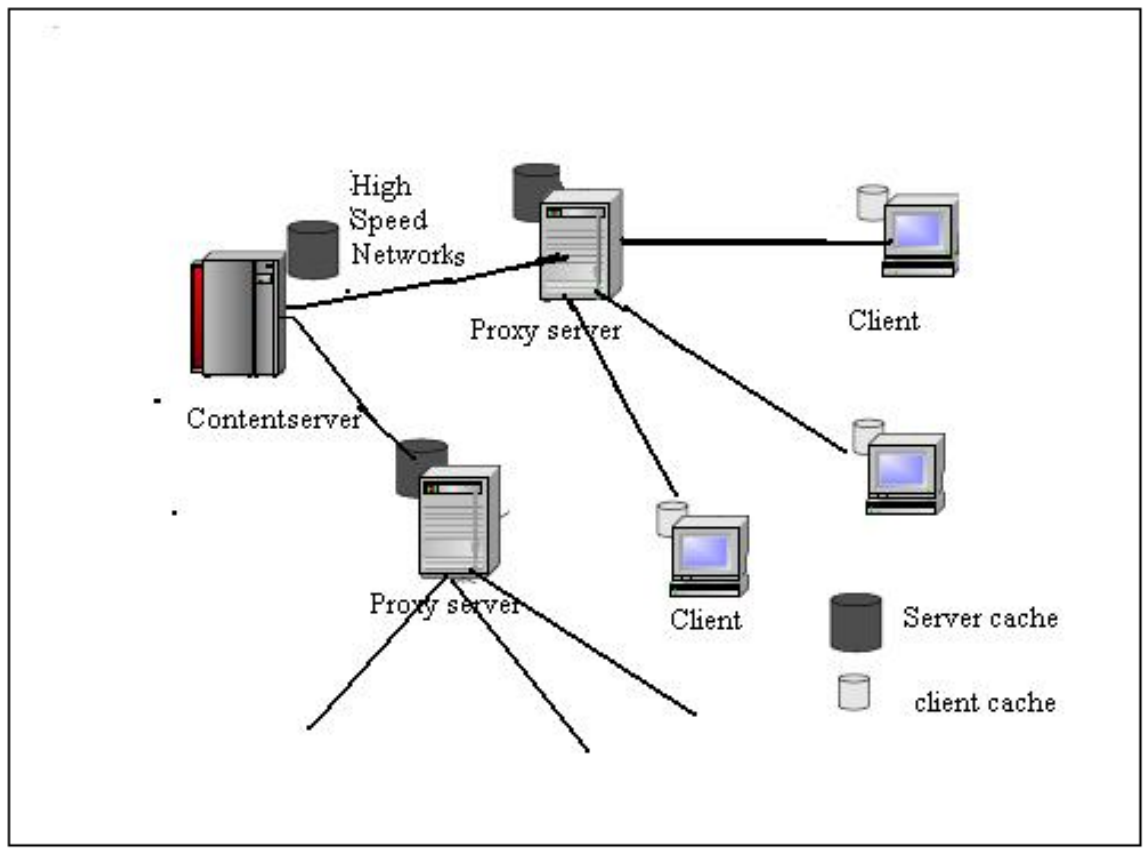

**Figure 1.** VOD architecture

### 2.2 Open-loop and closed-loop schemes

There are two types of VOD services: the “true” VOD and the “near” VOD. For the “true” VOD, clients are served immediately after their requests are received. Each customer is allocated by a transmission channel and a set of server resources. As the number of customer requests increases, the quality of the service can be maintained only by increasing server resources and network bandwidth, which ultimately leads to  expensive to operate, and non-scalable system.

Near-VoD (NVOD) makes use of multicast delivery to service more than one customer with a single set of resources to substantially reduce the system cost and achieve scalability. In this system, each movie can be multicast using a predetermined number of channels. For each channel, the assigned movie is repeated over and over and channels transmitting the same movie are offset by a time slot. Movies are available only at the beginning of these slots (say 15–30 min). A customer making a request after the start of a multicast channel will thus have to wait till the upcoming channel starts transmitting the movie. This introduces a significant start-up delay to the customer, which effectively contradicts the on demand nature of the service [8].

NVOD schemes may be classified in two categories: open-loop schemes and closed-loop schemes. In most closed-loop schemes, the server allocates channels and schedules transmission of video streams based on client requests using batching, or patching techniques. In batching, requests for the same video clip are delayed for a certain amount of time to serve as many requests as possible with one multicast channel. Open-loop schemes require constant bandwidth regardless of request rates. However, open-loop schemes are not adaptive because the server broadcasts a constant amount of video streams regardless of whether there is an outstanding request or not. Thus, while an open-loop approach can support an unlimited number of requests for popular video clips with a constant amount of bandwidth, it wastes bandwidth when the request frequency is low.

Closed-loop schemes permit some feedback that allows the server to adapt the transmission to the client requests. By storing frequently requested movies on cache of the servers, most client requests can be served .Minimizing the bandwidth and reducing access latencies are the two primary considerations in the design of multimedia streaming server architecture. Modern multimedia servers come with huge caches and some of these clips can easily be accommodated in part or fully in the cache. Also, while it might be very beneficial to store the video streams that are getting accessed in the cache completely, limitations on how much of each stream can be stored in the cache necessitates explicit caching mechanisms to store multimedia streams.

The challenge is to exploit the knowledge that there are many concurrent viewers of the hot movie in order to provide the service more efficiently. The solutions entail providing to an unlimited number of viewers of the same movie similar service flexibility to that of VOD at a reasonable cost to the server and communication network.

### 2.3 Multicast and Interval Caching

A client which requests for data which is in the cache can be served from there instead of being refused admission for lack of disk access time.

Whenever the bandwidth bottleneck limits the number of clients a multimedia streaming server can simultaneously support, a multicast delivery with interval caching [3][4] becomes particularly attractive for multimedia applications. Interval caching exploits the high skew ness in video access patterns by attempting to pair each playback request with an immediately preceding request for the same video that is currently being serviced from the cache. Specifically, interval caching reuses the data brought by a stream in servicing a closely following stream. The two streams are called the following stream and the preceding stream, respectively.

Only the initial portion of a media object, called the *prefix*, is streamed from the server's cache. Batching is an approach used to exploit the memory bandwidth and to save disk bandwidth in media servers by defining temporal cycles called batching windows. All requests that arrive within such a cycle are collected and at the end of the cycle, all requests to the same video are serviced from the same media object saved in the cache. Upon receiving a continuous request from the client, the server collects the entire request within a cycle and immediately delivers the prefix to the client at the end of the cycle. The basic approach is the creation of a multicast group for the delivery of a video stream to a requesting end-user. If another user requests the same video shortly after the start of this transmission, the request is considered as multicast transmission. This makes the cache sharable between as many clients as possible. Half portion of the cache is used to store only the prefix of the most popular videos and when deciding what needs to be removed from the cache during cache replacement it might be a good idea to retain initial portions of a file as opposed to the later portions since it could increase the chances of clients getting admitted. Thus, the blocks storing later portions of data for a multimedia stream can be marked as more eligible for replacement than the one storing only the prefix. Naturally, an increase in cache hit percentage would have a direct bearing on the number of clients that can be admitted and served without missing deadlines. Our approach thus balances the time duration of a batching window and deadline of each requested video.

Moreover, the number of initial segments cached is dynamically determined by the popularity of an object. It is assumed that the latency between a client device and the proxy server is negligibly small, but the latency between the proxy server and the content server is relatively large and cannot be ignored. Here, we assume that we are dealing with only clients that are reading from the disk and not writing to the disk.

As the cache is generally faster than the hard disk retrieval time directly from server, the startup delay for a playback of each block can be remarkably reduced even at the expense of batching time.

As an example, recent studies found that nearly 90 percent of media playbacks are terminated prematurely by clients after watching the initial portion of the video. Hence, it fetches the remaining portion, the *suffix*, from the server's disk to the prefetch cache relaying on the client's request.

## 3 RELATED WORKS

Admission control policies are typically divided into three classes: deterministic, predictive and statistical.

### 3.1 Deterministic Admission Control

A deterministic service policy [5] guarantees specified QOS requirements of existing customers and admits a new client only if its service demand does not affect the present clients, Since this strategy assumes the worst case cache performance, it provides 100 % service guarantee to the clients in the server but underutilizes the server capacity.

### 3.2 Predictive-Average Admission Control

A predictive admission control policy [6] monitors the server utilization over a time window and admits a client if the recent history indicates that the server can meet the requirements of the client. This admission control strategy results in higher server throughput than the deterministic admission control strategy, but needs to collect real-time history of the server for making decisions.

### 3.3 Statistical-Average Admission Control

A statistical admission control policy [2] uses the server load usage statistics to admit new customers. The last two admission policies provide soft QOS guarantee. By considering the fact that a deterministic policy is likely to underutilize the resources and a predictive admission control algorithm may incur high overhead to collect real-time workload history, statistical admission control schemes have become attractive for QOS assurance since they can guarantee service with acceptable QOS violations.

To improve the multimedia streaming services many studies on the caching of multimedia streaming objects have recently been studied. Among this Dan and Sitaram proposed a caching scheme for VOD servers named interval caching, which exploits the temporal locality of accessing the same multimedia object consecutively [7]. The interval caching (IC) scheme consist of all consecutive request pairs by the increasing order of memory space requirements, and then allocates memory space to as many of the consecutive pairs as possible. When an interval is cached, the following stream could read the data without any disk access since it will be directly served from the buffer cache.

Nachum et. al., proposed a Real-time multicast Communication admission control procedure [1], which consider requests of multiple streams from multiple destinations and resolve contention when users' requests exceed available network resources. However, the interval caching based on popularity and to retain the prefix, the initial portion of the video as opposed to later is not considered.

Ohhoon Kwon has proposed a popularity and prefix Aware Interval caching for multimedia streaming servers, but they have nit extended their work on the admission control and increasing the number of concurrent users by multicast transmission.

## 4 THE ARCHITECTURE OF MULTICAST VOD SYSTEM

In video delivery system the popularity and access pattern plays an important role. Because different videos are requested at different rates and at different times, videos are usually divided into most popular and less popular and requests for the top 10-20 videos are known to constitute 60-80% of the total demand.

So, it is crucial to improve the service efficiency of most popular videos. Thus, requests by multiple clients for the same video arriving within a short time interval can be batched together in the cache and serviced using a single stream. This is referred to as batching.

The Multicast facility of modern communication networks offers an efficient means of one-to-many data transmission. The basic idea is to avoid transmitting the same packet more than once. Since all the proxy servers are connected to the content servers, as the popularity of any video increases the number requests arriving from the proxies to the content server also increases.

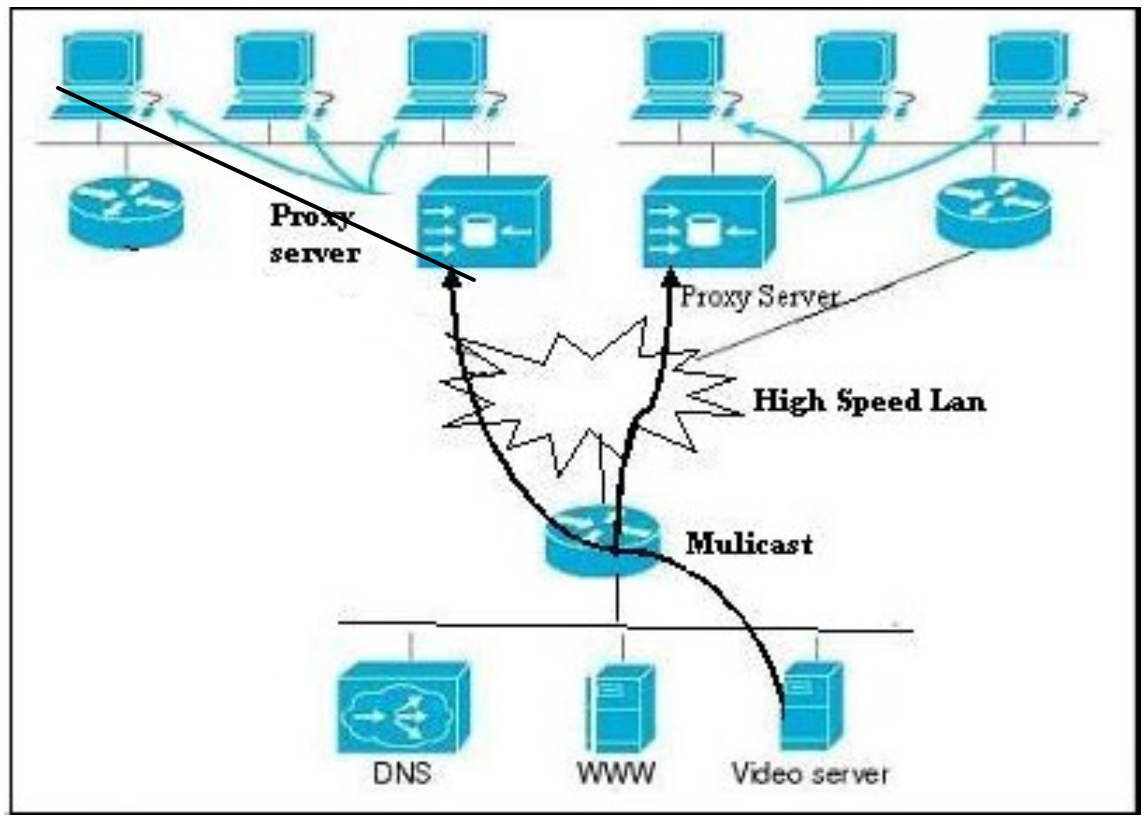

**Figure 2.** A multicasts VOD System

The same copy of prefix of the video can be multicast to all the proxies and sometimes directly to the clients by batching the entire request together with a small start up delay due to batching. This signicantly improve the VOD performance, because it reduces the required network bandwidth greatly, thereby decreasing the overall network load improves the system throughput and enables servicing a large number of clients and provides excellent cost/performance benefits.

In spite of these advantages, Multicast VOD has the following challenges viz,
It is difficult to maintain the VCR-like support. Batching makes the clients arriving at different times share a multicast stream, which may incur long service latency. The routers should support the multicast of VOD. Some of these challenges are overcome in this paper usually VCR-like support is not expected from the Proxy servers. The service latency problem is overcome by starting the service to the whole batch within the deadline constraints of the first request as shown in figure 3.

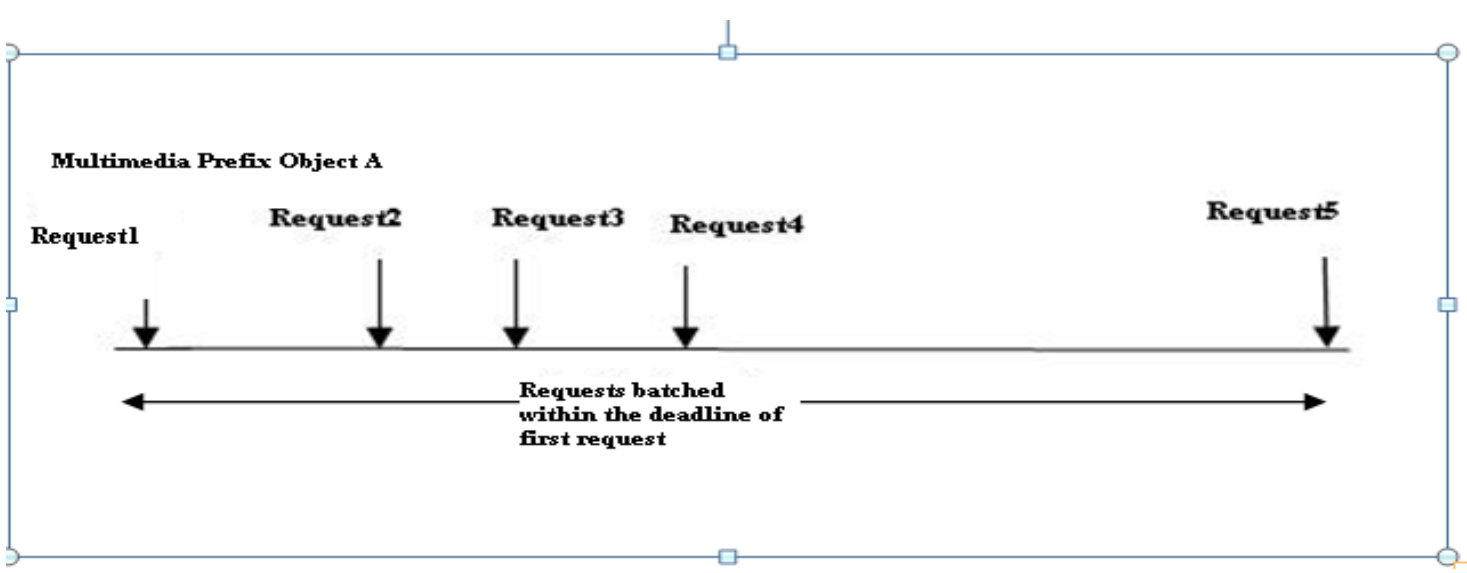

**Figure 3.** Batching of requests

## 5. PREFIX POPULARITY AWARE INTERVAL CACHING MULTICAST TRANSMISSION PROBLEM FORMULATION

A batching of requests for prefix is formed by having multiple clients to receive the same VoD application and is identified by a unique multicast address. A batch Bn is defined as a batch of users listening to the same multicast channel Cn. When a user A desires to join a particular prefix batch, for the prefix present in the cache, the service manager will provide the user with the batch's multicast address and its size. Before the actual reception of the requested video data, the user A will be asked to multicast a session packet to all the members of the batch it is willing to join.

The session packet and reply packets contain a source-ID and a timestamp. Assume that user A sends a session packet $P_s$ at time $t_1$ to the entire multicast group. The one group transmitting the prefix of the same video whose deadline constraints matches with that of the new request is admitted. The network bandwidth is updated.

The service manager will receive the information about the multicast group. Session profile will be updated by the manager using this information. Each session profile is identified by a name (i.e., movie's name). In addition to information about movie sequence statistics (i.e. frame rate) and users' buffer size, the session profile contains the following major elements:

- Client ID—defines a user/member of the batch.
- Start-time—defines the time a user started viewing a movie.
- Multicast-channel—defines the multicast channel a user is listening to.
- Establishment-time—defines the time an interconnection was established between two users.
- Expiration-time—defines the time a multicast group's admission time will expire.

To ensure the fact the video data must be continues, the server must employ an admission control algorithm, before accepting new request to a multicast group. First the adequate resources are available to the new request throughout the entire path from the video server to the proxy server. Second, the acceptance of the new request to the preexisting multicast session should not affect the performance requirements of other clients being already in service.

The server is responsible for servicing all the admitted clients in each round. But the server is limited by the resources, and it should not get overloaded. A server rejects client, in case if found overloaded. Caching reduces the retrieval time considerably also adds more requests being streamed. Retrieval time of the hard disk is dependent on the rotational time, seek time and transfer time. Compared with the unicast transmission from the cache, it is more beneficial when multiple clients are requesting for the same video, with a slight delay between each other.

Caches can be placed directly in front of a particular server, to reduce the number of requests that the server must handle. Most proxy caches can be used in this fashion, but this form has a different name (reverse cache, inverse cache, or sometimes http accelerator) to reflect the fact that it caches objects for many clients but from (usually) only one server.

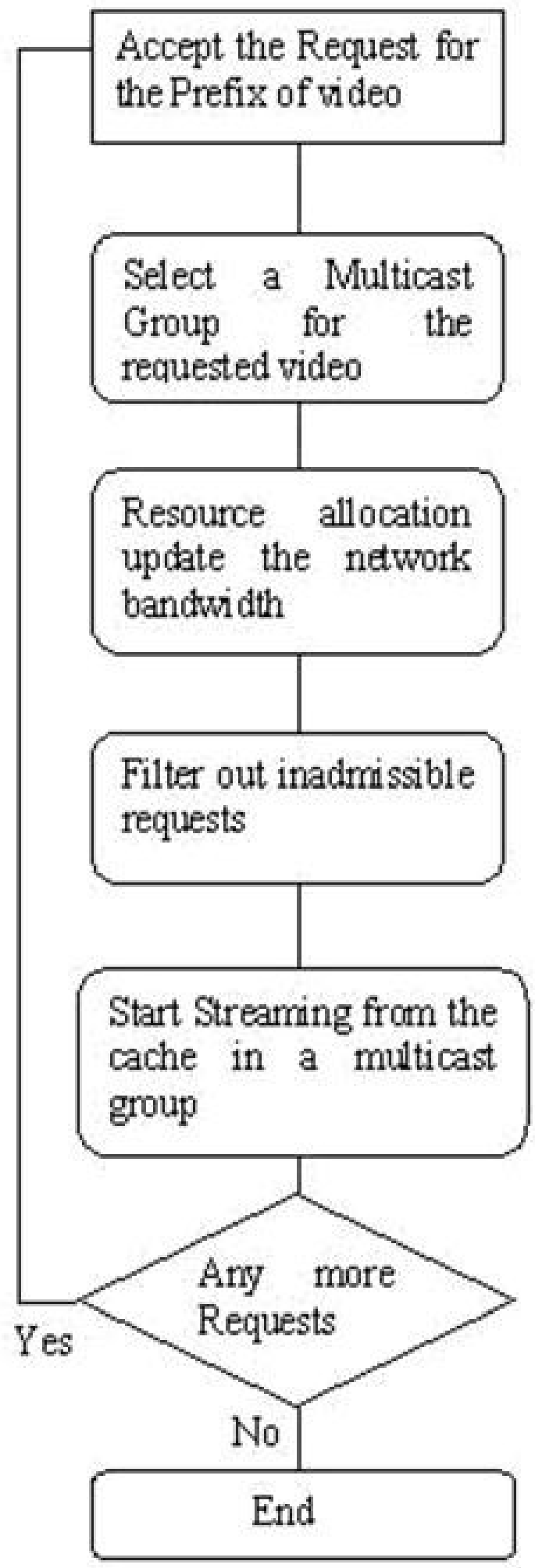

**Figure 4.** General flow diagram for the Admission Process

**Table 1**: Parameters in the model

| | |
|---|---|
| Db | Disk Bandwidth |
| Cb | Cache bandwidth |
| Nb | Network Bandwidth |
| $S_1,S_2, S_3,S_4\ldots$ | Strands in a video |
| $f_1,f_2, f_3\ldots$ | Number of frames from each Strand |
| S | Total length of a video |
| Frame Size | F *bytes*.. |
| *T* | Service Time |
| $r_i$ | Bit rate of a client |
| *R*. | defined as the minimum of the |

Let us assume that a client *i* needs a multimedia session. Each client *i* is retrieving a video strand say $S_1,S_2, S_3,S_4\ldots$respectively for any video stored in the hard disk. Let fi be the number of frames of data for a stream *i* in the cache. The total number of frames that must be retrieved is specified as the service requirement of any client. Let $f_1,f_2, f_3\ldots$ denote the number of frames that must be retrieved as a percentage from the strands $S_1,S_2, S_3,S_4\ldots$respectively during each round

$P_R^i$ denotes the playback rate of ith client which is frames/sec.

Each of these clients can have different playback rates and therefore needs the disk scheduler to return blocks to them at different rates. Let the frame size be of F *bytes*. Multimedia clients have periodic access to the disk, since the server has to serve the frames at regular intervals .The duration of the interval is decided by the data rate that the client is requesting. It is usually dependent on the highest data rate that has been requested by the admitted clients**.** A client with a data rate of 20 *Kbps* will have to be served twice as often as a 10 *Kbps* client in a given time.

Therefore, it is very important to decide on the duration of a round.

Let the time in a round be denoted by *R*. This duration *R*. is decided as follows:

$P_R^i$ being playback rate (expressed in terms of frames/sec) of strand Si the duration of a round $\mathcal{R}$ , defined as the minimum of the playback durations of the frames accessed during a round, is given by a round, is given by

$$\mathcal{R} = \min_{i\,\varepsilon[1,n]} \frac{fi}{P_R^i}$$

Service time $\mathcal{T}$ the total time needed in retrieving frames from disk during each round should not exceed $\mathcal{R}$ ,to ensure the continuous playback of each media stream. Cache manager can obtain information on the size of the stream from the file system.

The one and only criterion that is used to determine whether a client requesting a multimedia session can be admitted is the availability of disk bandwidth to satisfy the QoS requirements specified by the client. Let there be $n$ clients in the system, when a new client $(n + 1)$ requests admission. Let the total size of the stream (file) be $Fi$ frames . Further, let the bit rate of the client be, as in the, $r_i$ *bps*, the time taken for playing the whole stream will be rate of the client be, as in the, $r_i$ *bps*, the time taken for playing the whole stream will be

$$t_i = \frac{F_i}{r_i}$$

In presence of cache data the total time taken to read the entire stream

$$t_i = \frac{F_i - C_i}{r_i}$$

From the relation it is clear that the total time taken to stream the entire video with reduce with the size $c_i$ which denotes the total number of blocks of data for a stream *i* in the cache. The new reduced bit-rate of the client would be

$$r_i' = \frac{Fi - Ci}{Fi} r_i$$

The residual bandwidth of any disk meets the minimum quality requirement of the new client which dependent on rotational latency and seek time, if it satisfies the following relation

$$\sum_{i=1}^{n} Fi * \left( T_{avg}^{seek} + T_{avg}^{rot} \right) <= \alpha * \mathcal{R}$$

Where α represents the fraction of bandwidth provided by the disk for the purpose of servicing multimedia requests only.

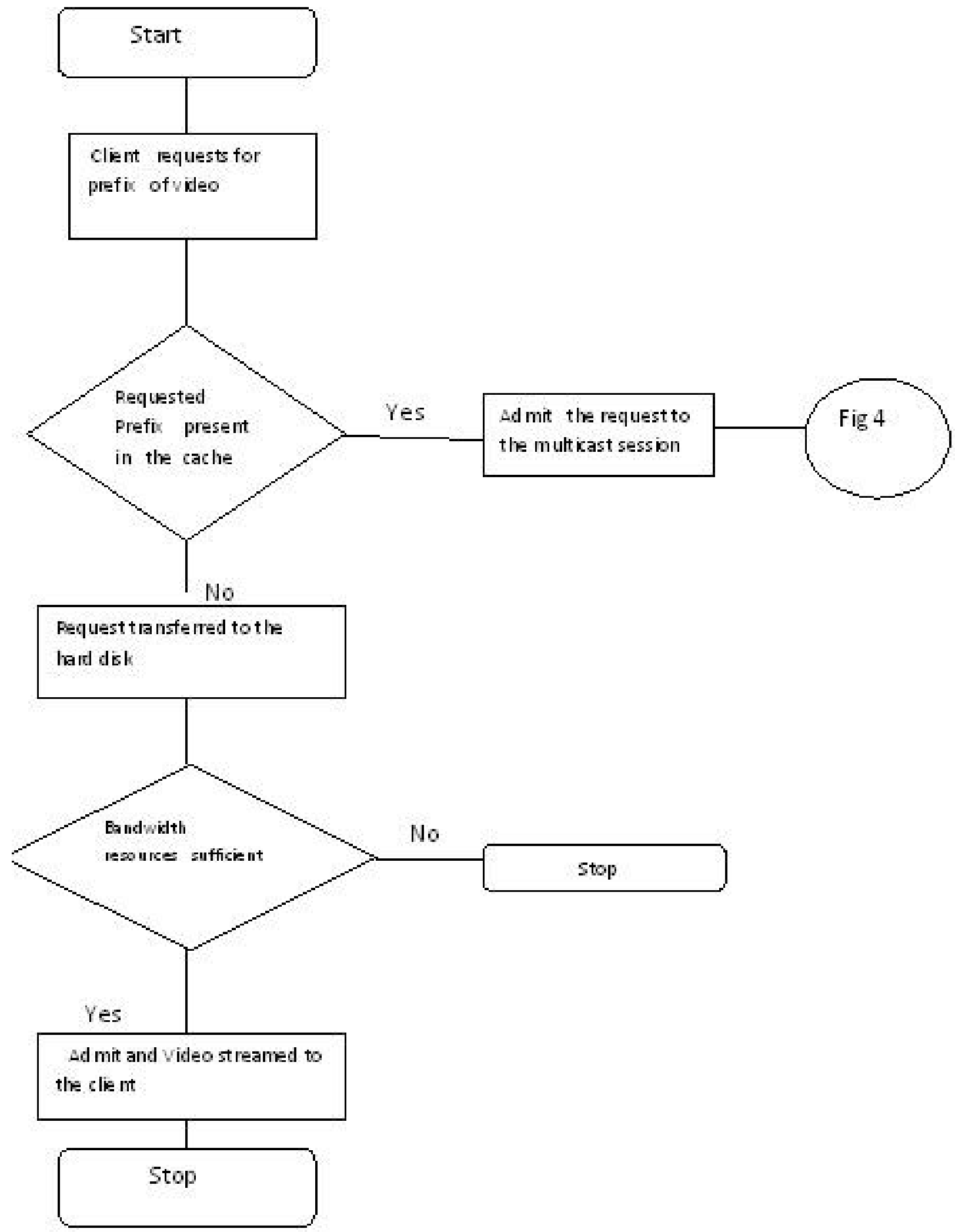
Start
Client requests for prefix of video
Requested Prefix present in the cache
Yes
Admit the request to the multicast session
Fig 4
No
Request transferred to the hard disk
Bandwidth resources sufficient
No
Stop
Yes
Admit and Video streamed to the client
Stop

## 6. THE PROPOSED MODEL FOR SIMULATION

The model is based on the following concepts:

1. A hard disk with a bandwidth of Db is supported with a cache of bandwidth Cb and Nb denotes maximum network bandwidth of high speed networks connected between the proxy server and the content server.
2. Total number of concurrent users serviced from the content server's disk is limited by its seeking time and rotational time. This adds the start up latency for each request resulting in under utilization of network and disk bandwidth. .
3. Cache can be modeled as another server. Assuming that 50% of cache space is used to store Prefix of most popular videos, with the arrival rate to the cache, $\lambda c$ and the average time spent by a user in the cache, $T_c$ (service time of the cache), are known, the average number of concurrent users in the cache, *Nc,* can be calculated by Little's law ***(Nc*** = $\lambda c$ * $T_c$*)*. This is limited by the network bandwidth and the deadline of the first request in the video.
4. For a given prefix (strand $S_1$) of the video stored in the server's cache, the server is allowed to batch clients requesting the same video for the first time and to serve clients in the same batch with one multicast video stream. This approach has the advantage that it can save server resources as well as server access and network bandwidth, thus allowing the server to handle a large number of clients without sacrificing access latency.
5. The subsequent frames of the strands of the videos (the suffix) are streamed later streamed from the cache or from the disk depending on the popularity growth rate for a video.
6. While replacing the videos on the cache, the blocks storing later portions are considered more eligible for replacement than one storing only the prefix, increasing the hit ratio for the first time requested videos.

## 7. SIMULATION MODEL

Our simulation model consists of a central server with 100 complete movies stored in it. A cache has the space to hold 10 movies. But we have assumed that a movie is dividing into four strands$S_1$, S2, S3, S4.

$S_1$ denotes the prefix of the video. 50% of the cache is used to store only the prefix and remaining holds the later parts of the movie being streamed Central content server is connected to the proxy servers with high speed networks (ATM).Total start up latency of the hard disk is appox 6 ms. Mean number of blocks per video is 200, mean inter arrival time is 60 s. Maximum hard disk bandwidth 10Mbytes/sec, client requests at a bit rate of 300Kbits/sec.

## 8. SIMULATION RESULTS

The simulation results are as presented below.

Fig 6 proves that total number of videos streamed is considerably more in our model when compared with Statistical and Deterministic. Fig 7 demonstrates that the total videos rejected while getting admitted is less when more priority is for prefix when compared with only popularity aware interval caching. Fig 8 shows our model utilizes the hard disk more efficiently. Fig 9 shows that by an efficient replacement policy we can increase the Hit ratio.

## 9. CONCLUSION

In this paper we have shown that by storing only the prefix of most popular video in the cache and batching all the requests with the time interval of deadline of first request and then multicasting the video to all the requested clients we provide admission for more clients. This is compared with the standard admission control approaches such as deterministic and statistical admission control policy.
The prefix definitely increases the hit ratio and reduces the rejection ratio. The future work is carried out on an efficient replacement technique and to dynamically allocate the blocks to the frames in the cache depending on popularity.

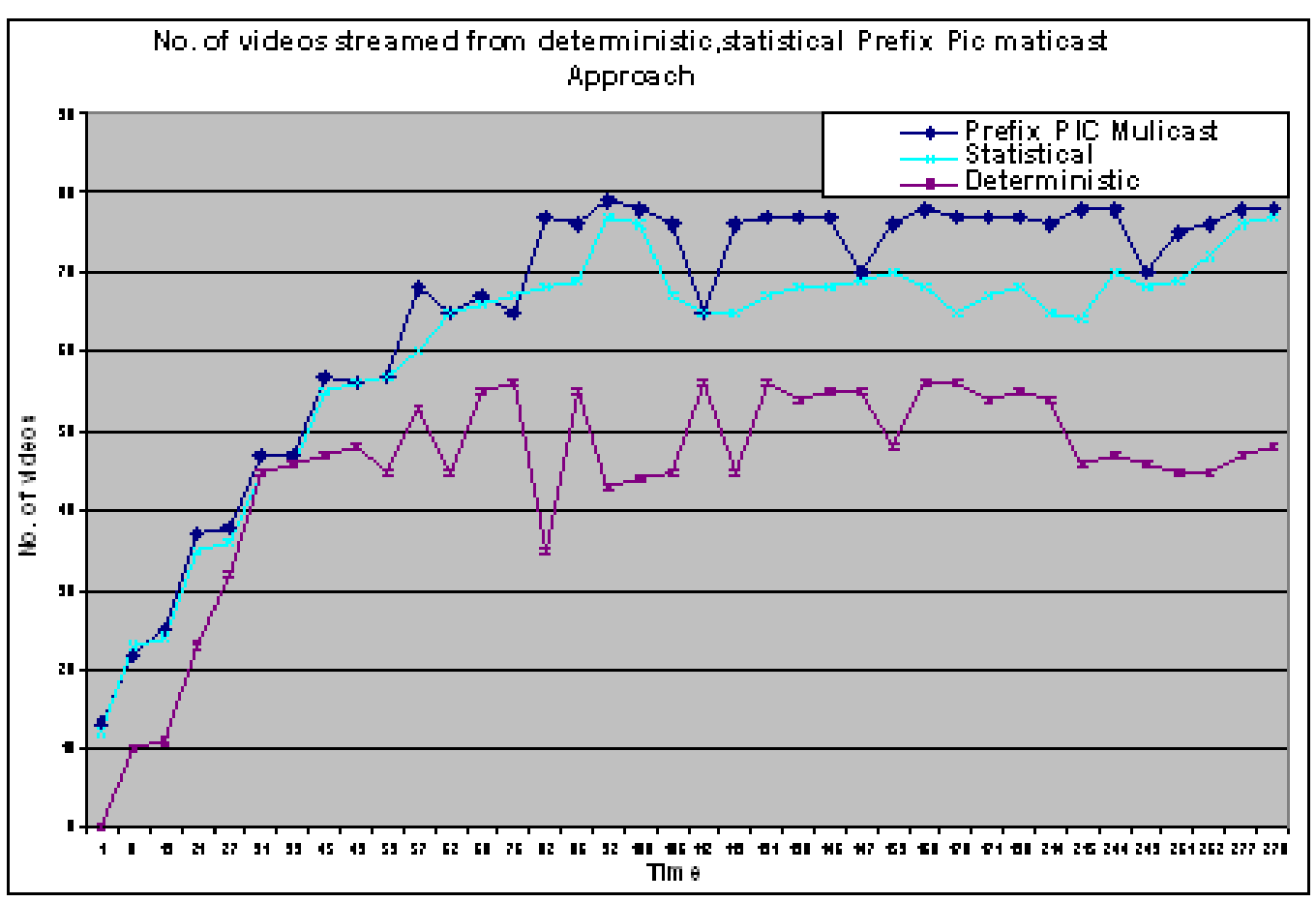

**Figure 6.** Comparison between the total number of videos streamed from Deterministic, Statistical, Prefix PIC multicast Approach

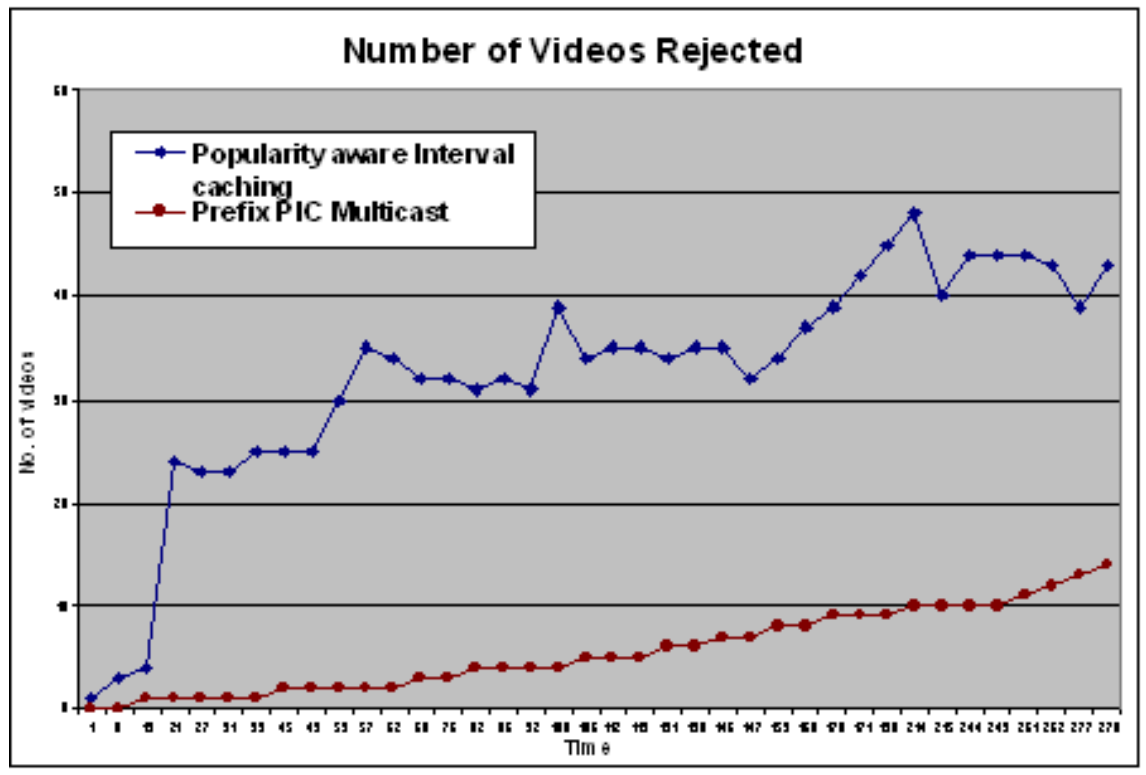

**Figure 7.** Total number of videos rejected and getting admitted when prefix with PIC is given more priority compared with only PIC.

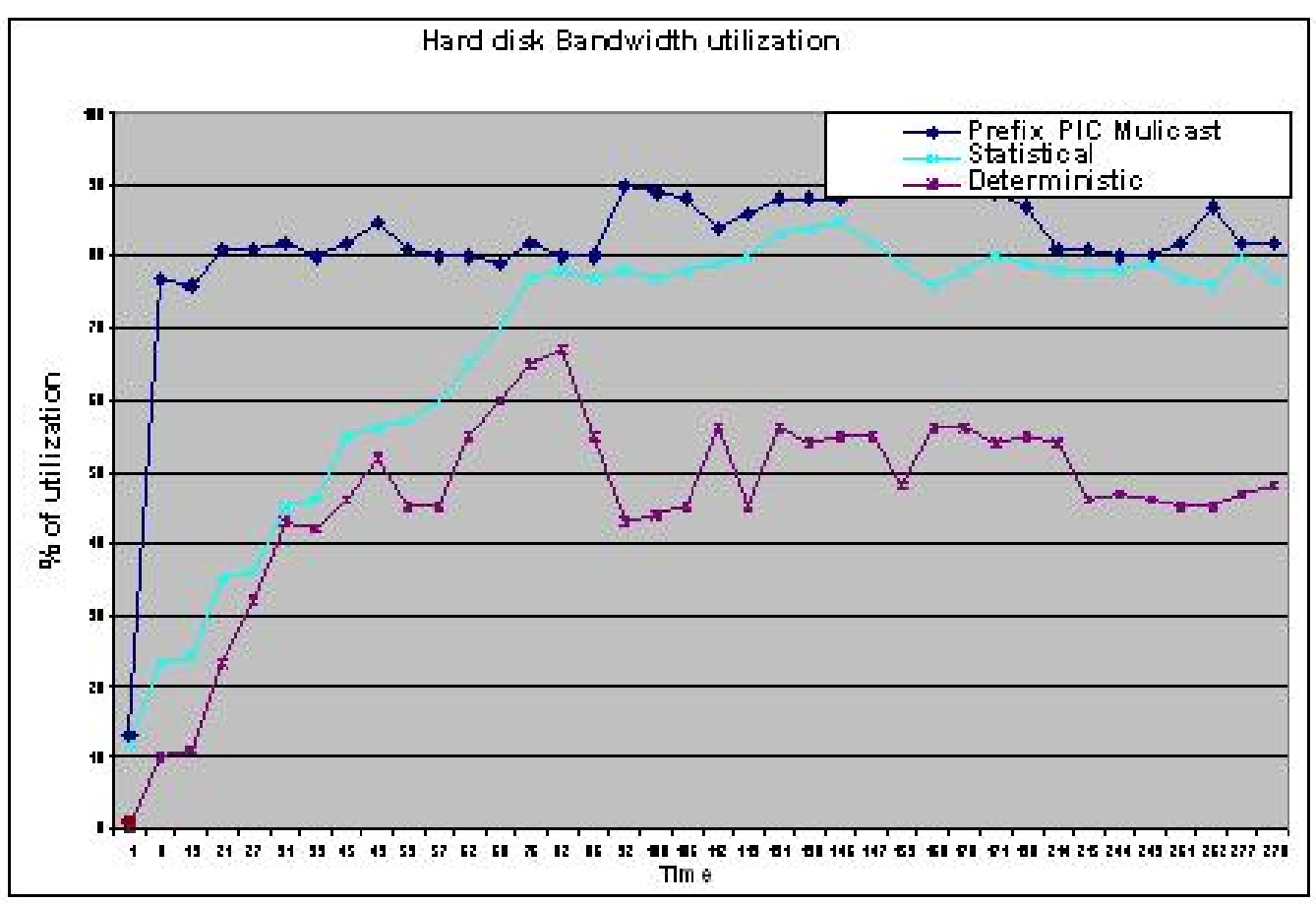

**Figure 8.** Percentage of Hard disk bandwidth utilization

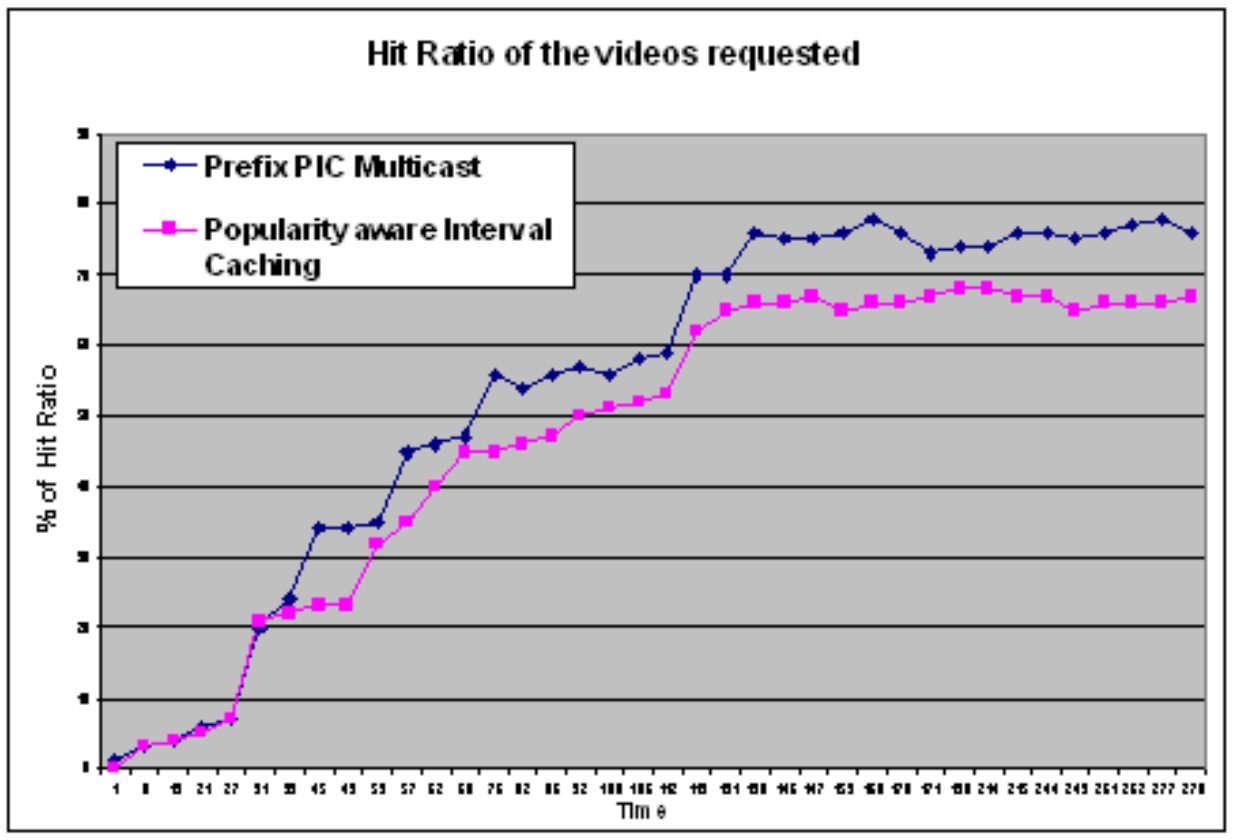

**Figure 9.** Percentage of Hit ratios of the videos requested when prefix with PIC is given more priority compared with only PIC.

## REFRENCES

[1] Nachum Shacham, *Fellow, IEEE* and Johannes Dengler, Christoph Bernhardt "Deterministic admission control strategies in video servers with variable bit rate streams Interactive Distributed Multimedia systems and Services "Volume 1045/1996, pp 245-264

[2] Harrick M. Vin, Pawan Goyal, Alok Goyal, and Anshuman Goyal "A Statistical Admission Control Algorithm for Multimedia Servers"

In Proceedings of the ACM International Conference on Multimedia Multimedia'94, San Francisco, October 1994

[3] Ohhoon Kwon Hyokyung Bahn Kern Koh "Popularity and Prefix Aware Interval Caching for Multimedia Streaming Servers**"** Computer and Information Technology, 2008. CIT 2008, 8th IEEE International Conference 8-11july2008, pp 555-560

[4] Sun-Euy Kim and Chita R. Das **"**A Reliable Statistical Admission Control Strategy for Interactive Video-On-Demand Servers with Interval Caching" Proceedings of the Proceedings of the 2000 International Conference on Parallel Processing, pp 135

[5] Johannes Dengler, Christoph Bernhardt "Deterministic admission control strategies in video servers with variable bit rate streams Interactive Distributed Multimedia systems and Services "Volume 1045/1996, pp 245-264

[6] Vin H, Goyal A, Goyal Al, Goyal P (1994) "An observation-based admission control algorithm for multimedia servers." Proceedings of the IEEE International Conference on Multimedia Systems and Computing, Boston, IEEE Computer Society Press, Los Alamitos, pp 234–243

[7] A. Dan and D. Sitaram, "A Generalized Interval Caching Policy for Mixed Interactive and Long Video Environments," SPIE Multimedia Computing and Networking Conf., San Jose, CA, 1996

[8] C.C. Aggarwal, J.L. Wolf, The maximum factor queue length batching scheme for video-on-demand systems, IEEE Tran. Comput. 50 (2) (February 2001) pp 97–110